\documentclass[a4paper,twoside]{article}
\usepackage{amssymb,amsmath,lscape}
\usepackage[authoryear]{natbib}
\bibpunct{(}{)}{;}{a}{}{,}
\usepackage{graphicx}
\date{} %Please leave the date blank
\title{\large\bf\flushleft Metallicity distributions in and around galaxies}
\author{\parbox{\textwidth}{\flushleft
\vspace{-0.5cm}
{\it Gabriella De Lucia}\\
\vspace{0.4cm}
{\small \,INAF-Astronomical Observatory of Trieste, via G.B. Tiepolo
  11, I-34143 Trieste, Italy \\ Email: delucia@oats.inaf.it}}}
\begin{document}
\twocolumn[
\begin{changemargin}{.8cm}{.5cm}
\begin{minipage}{.9\textwidth}
\vspace{-1cm}
\maketitle
%
%
%%%%%%%%%%%%%     ABSTRACT    %%%%%%%%%%%%%
%Abstract of no more than 200 words here.
\small{\bf Abstract:} Metals are found in all baryonic phases and environments,
and our knowledge of their distribution `in and around galaxies' has
significantly improved over the past few years. Theoretical work has shown that
the fraction of metals in different baryonic components can vary significantly
when different feedback schemes are adopted.  Therefore, studies of element
abundances provide important information about all gas-dynamical processes
which determine the cosmic evolution of baryons.  I give here a brief review of
recent observational progress, describe the implications of recent theoretical
studies, and discuss briefly future prospects.

%%%%%%%%%%%%%     KEYWORDS    %%%%%%%%%%%%%
\medskip{\bf Keywords:} stars: abundances --- ISM: abundances --- galaxies:
abundances -- galaxies: evolution -- galaxies: high-redshift
% Please write all keywords in lower case. PASA uses the
% standard list of subject headings adopted by The Astrophysical Journal
% and available from http://www.journals.uchicago.edu/ApJ/keywords_text.html.
% Keywords are separated by em-dashes, i.e. ---

%%%%%%%%DO NOT EDIT%%%%%%%%%%%%
\medskip
\medskip
\end{minipage}
\end{changemargin}
]
\small
%%%%%%%%EDIT FROM HERE%%%%%%%%%%%%

\section{Introduction}

During the last decade, significant progress has been made in the study of
element abundances. Accurate measurements of metallicities have been collected
for a large number of individual stars for our own Galaxy, and for a number of
the brightest members of the Local Group. Large scale surveys such as the 2dF
galaxy redshift survey and the Sloan Digital Sky Survey (SDSS) have allowed
abundance determinations in hundreds of nearby galaxies. The new generation of
large telescopes and sensitive detectors have allowed us to measure a range of
chemical elements in stars, HII regions, cold interstellar gas, and hot
intergalactic medium, up to an epoch when the Universe was only about one tenth
of its current age.

As I will show later, the distribution and amount of heavy elements in the
different bayonic components of the Universe, is sensitive to physical
mechanisms that regulate the evolution of bayons as a function of cosmic
time. Therefore, the main motivation of chemical abundance studies is the
possibility to use this information as a tool to reconstruct the past history
of star formation, and the relative role of feedback processes and various
gas-dynamical processes in determining the observed cosmic evolution of baryons
(for a classical review on chemical evolution, see \citealt{Tinsley_1980}; for
a more recent review focused on the Milky Way and its satellites, see
\citealt{Matteucci_2008}).

As mentioned above, metals are present in all baryonic phases and
`environments'. Even in rather low density regions, the inter-galactic medium
appears to be polluted to $\gtrsim 1$ per cent solar metallicity, as measured
by metal lines in the Lyman alpha absorption systems (see section
\ref{sec:igm}). How and when this chemical enrichment occurred exactly, is
still matter of debate. From the theoretical point of view, various physical
processes can provide viable explanations for the transfer of metals from the
stars in galaxies to the inter-galactic and intra-cluster medium, for example
ejection of enriched material during mergers, tidal/ram-pressure stripping,
galactic outflows (e.g. supernova driven outflows). The relative importance of
these various channels, and the details of how metals are transferred from the
stellar to the gaseous (cold and/or hot) phase remain to bee understood.

Various techniques have to be used to measure the metal abundances in different
baryonic components. Stellar population synthesis models have been extensively
used for nearby galaxies (in particular for elliptical galaxies and, in a few
cases where the contamination of stellar absorption features by nebular
emission can be removed, for late-type galaxies). Optical nebular emission
lines have been used to measure the gas phase metallicity of star forming
galaxies up to $z\sim 1$. The same method has been used for Lyman break and
UV-selected galaxies up to $z\sim 3$. X-ray spectroscopy is necessary to
measure the metal content of the intra-cluster medium (ICM). As I will mention
in the following, each of these techniques has its own problems and limits, and
at least some of them are far from being firmly calibrated. 

Chemical enrichment studies have a quite long history that cannot be summarised
in a few pages. Given the vastity of the subject, I have deliberately decided
to focus mainly on recent observational results and theoretical developments,
rather than providing an historical review. In the next sections, in
particular, I will provide a brief overview of what is known, observationally,
about the metallicity distributions in our own Galaxy (focusing on the stellar
halo), in the local Universe and in the ICM and IGM. I will then briefly touch
upon the evolution of the mass-metallicity relation and metallicity
measurements at higher redshift. In each section, I will mention recent
theoretical progress, and I will conclude with a few remarks and considerations
about future prospects.

\section{Metallicity distributions in galaxies - the stellar halo and the
  satellites of our Milky Way.}  

Our own galaxy - the Milky Way - is a fairly large spiral galaxy consisting of
four main stellar components. Most of the stars are distributed in a thin disk,
exhibit a wide range of ages, and are on high angular momentum orbits. A much
smaller mass of stars (about 10-20 per cent of that in the thin disk) reside in
a distinct component which is referred to as the `thick disk'. The stars in the
thick disk are old, have on average lower metallicity than those of similar age
in the thin disk, and are on orbits of lower angular momentum. The Galactic
bulge is dominated by an old and relatively metal-rich stellar population with
a tail to low abundances. The fourth component - the stellar halo - represents
only a tiny fraction of the total stellar mass ($\sim 2\times10^9\,{\rm
  M}_{\odot}$) and is dominated by old and metal poor stars which reside on low
angular momentum orbits \citep{Freeman_Bland-Hawthorn_2002}.

Accurate measurements of metallicity, age, and kinematics have been collected
for a large number of individual stars in our own Galaxy. Historically,
chemical and kinematic information provided the basis for the first galaxy
formation models. \citet{Eggen_LyndenBell_Sandage_1962} studied a sample of 221
dwarf stars and found that those with lowest metal abundance were moving on
highly elliptical orbits and had small angular momenta. The data were
interpreted as evidence that the oldest stars in the galaxy were formed out of
gas collapsing from the halo onto the plane of the galaxy, on a relatively
short time-scales (a few times $10^8$~years). About one decade later,
\citet{Searle_Zinn_1978}, found no radial abundance gradient in a sample of 177
red giants and 19 globular clusters. These observations led Searle \& Zinn to
formulate the hypothesis that the stellar halo (in particular the outer region
of the halo) formed through the agglomeration of many subgalactic fragments,
that may be similar to the surviving dwarf spheroidal satellites (dSphs) of the
Milky Way.

\begin{figure}[ht]
\begin{center}
\includegraphics[scale=0.38, angle=0]{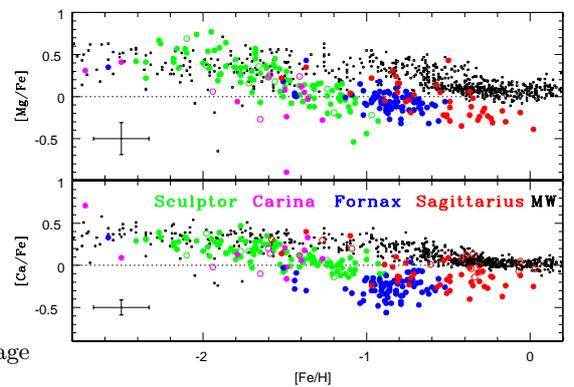}
\caption{From \citet{Tolstoy_etal_2009}. $\alpha$ elements (Mg and Ca) in four
  nearby dwarf spheroidal galaxies (coloured symbols). The small black symbols
  are a compilation of Milky Way disk and halo star abundances (see original
  paper for details).}
\label{fig:tolstoy}
\end{center}
\end{figure}

Although the Searle \& Zinn scenario appears to be in qualitative agreement with
expectations from the hierarchical cold dark matter scenario, the debate
between a rapid collapse and a sequence of accretion events is not closed. One
problem with the Searle \& Zinn scenario was pointed out by
\citet{Shetrone_Cote_Sargent_2001} who obtained high resolution spectra for 17
stars in three dSph galaxies and noted that stars in the Local Group dSphs tend
to have lower alpha abundances than stars in the stellar halo. These results
have been later confirmed with larger samples of stars:
e.g. Fig~\ref{fig:tolstoy} shows the abundance of magnesium and calcium in four
nearby dwarf spheroidal galaxies, compared with a compilation of Milky Way and
halo star abundances. These observations suggest that
the Galactic stellar halo cannot result from the disruption of satellite
galaxies similar to those studied by Shetrone and collaborators or Tolstoy and
collaborators.  It is, however, not entirely surprising that the {\it
  surviving} satellites might be intrinsically different from the main
contributors to the stellar halo. The argument has been put forward in a recent
study by \citet{Font_etal_2006} who reproduced the observed chemical abundance
pattern by combining mass accretion histories of galaxy-size haloes with a
chemical evolution model for individual satellites. In the model proposed by
Font and collaborators, the main contributors to the stellar halo were accreted
early on and were enriched in $\alpha$ elements by Type II supernovae, while
the surviving satellites were accreted later and have more extended star
formation histories and stellar populations enriched to solar level by both
Type II and Type Ia supernovae (see also \citealt{DeLucia_Helmi_2008}).

\begin{figure}[ht]
\begin{center}
\includegraphics[scale=0.42, angle=0]{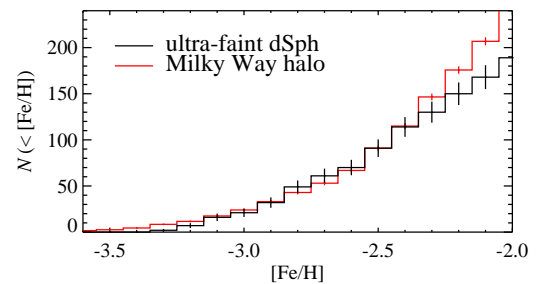}
\caption{From \citet{Kirby_etal_2008}. Cumulative metallicity distribution
  function for the metal-poor tails of the eight ultra-faint dSphs (black) and
  the Milky Way halo (red). See original paper for more details.}
\label{fig:kirby}
\end{center}
\end{figure}

A potentially more serious problem with the Searle \& Zinn scenario was pointed
out by \citet{Helmi_etal_2006} who found a significant difference between the
metal-poor tail of the dSph metallicity distribution and that of the Galactic
halo. In a more recent study, however, \citet{Kirby_etal_2008} have presented
metallicity measurements for 298 individual red giant branch stars in eight of
the least luminous dSphs of the Milky Way, and have shown that the metallicity
distribution of their stars is similar to that of the stellar halo at the
metal-poor end (see Fig.~\ref{fig:kirby}). 

Detailed abundance data for many elements are now being collected for larger
samples of stars in the Milky Way and in several dSph galaxies. It is important
to note that each galaxy shows a wide spread in metallicity and that, in some
cases, there are clear abundance gradients. It remains to be understood if more
luminous dSphs (like those studied by Helmi and collaborators) also host
extremely metal-poor stars. Ongoing spectroscopic surveys and detailed
comparisons with cosmologically motivated models including star formation and
chemical enrichment, will provide in the next future crucial information on the
role of dSphs in building the stellar halo of our own Milky Way.

\section{Metallicity distributions in galaxies - the local Universe}

For galaxies outside our Local Group, metallicity measurements of individual
stars are unfeasible, and only integrated values can be derived. Estimates of
the ages and metallicities of the stellar population of nearby galaxies can be
inferred from the integrated spectra, using stellar population synthesis
models. These studies are, however, limited by the difficulty of
deriving independent constraints on the age, star formation history,
metallicity and dust content of a galaxy. These degeneracies can be broken, at
some level, using spectral diagnostics that are not sensitive to dust and have
different sensitivities to age and metallicity. Studies in this area have
focused, in particular, on a set of absorption features defined and calibrated
using the spectra of nearby stars observed at the Lick Observatory - the so
called Lick Indexes \citep{Faber_1973, Worthey_etal_1994}. These studies are
generally limited to early type galaxies because of the lack of hot stars in
the Lick library. It is important to note, however, that the age-metallicity
degeneracy can be broken only in a relative way, and that the absolute values of
the ages and metallicities derived depend strongly on the specific choice of
metal indices.

\begin{figure}[ht]
\begin{center}
\includegraphics[scale=0.4, angle=0]{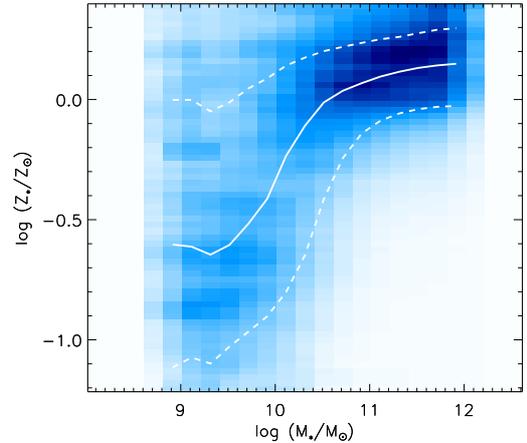}
\caption{From \citet{Gallazzi_etal_2005}. Conditional distribution of stellar
  metallicity as a function of the galaxy mass. The solid line indicates the
  median of the distribution and the dashed lines the 16th and 84th
  percentiles. See original paper for details.}
\label{fig:fig8_gallazzi}
\end{center}
\end{figure}

\begin{figure}[ht]
\begin{center}
\includegraphics[scale=0.4, angle=0]{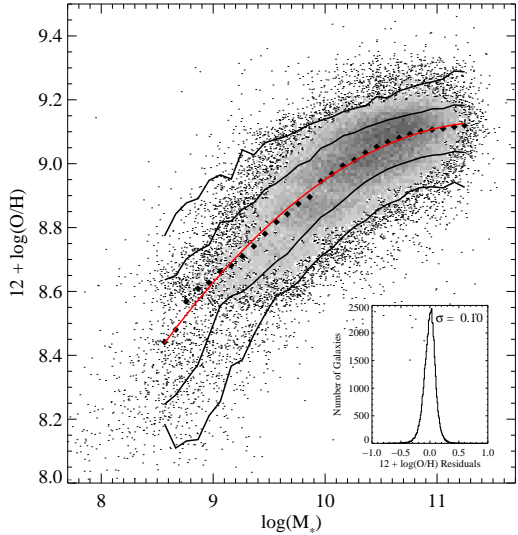}
\caption{From \citet{Tremonti_etal_2004}. Relation between stellar mass and
  gas-phase oxygen abundance for $\sim 5\times 10^4$ star forming galaxies in
  the SDSS. The large black symbols represent the median in bins of 0.1~dex in
  mass that include at least 100 data points. The solid lines mark the contours
  that enclose 68 and 95 per cent of the data. The red line is a polynomial fit
  to the data. The inset plot shows the residual to the fit. See original paper
  for details.}
\label{fig:fig6_tremonti}
\end{center}
\end{figure}

In a recent study, \citet{Gallazzi_etal_2005} have combined modern population
synthesis techniques with a Bayesian statistical approach to derive likelihood
distributions of ages and metallicities for a sample of $\sim 2\times10^5$
galaxies from the SDSS. The spectral features used in this study have been
selected to depend only weakly on the $\alpha$/Fe ratio, and estimates of ages
and metallicities have been determined also for star forming galaxies, by
removing the contamination of stellar absorption features by nebular emission.
Fig.~\ref{fig:fig8_gallazzi} shows the derived distribution of metallicity as a
function of the galaxy stellar mass, obtained by co-adding the two-dimensional
likelihood distributions for all galaxies. The figure shows that metallicity
increases with stellar mass. The 68 per cent confidence levels, however, show a
quite broad distribution, with a large (larger than the intrinsic
uncertainties) scatter for a given stellar mass.

A similar relation was observed between the gas-phase metallicity and the
stellar mass by \citet{Tremonti_etal_2004}, and is reproduced in
Fig.~\ref{fig:fig6_tremonti}.  Gas phase metallicities are usually estimated
using strong nebular emission lines, and calibrations that are based on
metallicity sensitive emission line ratios. Fig.~\ref{fig:fig6_tremonti} shows
the vast majority of galaxies having super-solar
metallicity\footnote{\citealt{Tremonti_etal_2004} adopt 12+log[O/H]=8.69 as
  solar value.}. It should be noted, however, that strong-line methods (as used
in Tremonti et al.) may systematically overestimate oxygen abundances by as
much as 0.2-0.5~dex \citep{Kennicutt_Bresolin_Garnett_2003}. Different
calibrations have been adopted in the literature: methods based on
photoionization models, empirical methods based on the measurements of the
electron temperature of the gas, or a combination of these. Unfortunately,
estimates obtained using different calibrations manifest large discrepancies,
and can produce mass-metallicity relations with different shapes, y-axis
offsets, and scatter (see \citealt{Kewley_Ellison_2008}). This complicates
comparisons between different samples and between samples at different
redshifts, for which different suites of emission lines are available.

Fig.~\ref{fig:fig6_tremonti} shows a rather linear correlation up to $\sim
10^{10.5}\,M_{\odot}$, after which a gradual flattening occurs. The correlation
is very tight, tighter than the one between stellar mass and stellar
metallicity, and extends over three decades in stellar mass. Tremonti and
collaborators interpret the shape of the mass-metallicity relation in terms of
efficient galactic outflows that remove metals from galaxies with shallow
potential wells, an idea that was originally introduced by
\citet{Larson_1974}. Alternative explanations have been proposed: e.g. a lower
star formation efficiency in low-mass galaxies \citep{Calura_etal_2009}, or a
variable integrated initial mass function \citep{Koeppen_Weidner_Kroupa_2007}.

There is a striking similarity between the two relations shown in
Figs.~\ref{fig:fig8_gallazzi} and \ref{fig:fig6_tremonti}. Gallazzi and
collaborators have compared their stellar metallicity measurements with
measurements of gas-phase metallicity, for those galaxies for which both
measurements are available. They have shown that there is a clear correlation
between these quantities, with approximately unit slope, but there is a large
scatter in stellar metallicity at fixed gas-phase metallicity. In a closed box
model, these two quantities should be tightly related. Therefore, these
findings indicate that gas ejection and/or accretion play an important role in
galaxy chemical evolution, and that their influence likely varies with stellar
mass.

\begin{figure}[ht]
\begin{center}
\includegraphics[scale=0.36, angle=0]{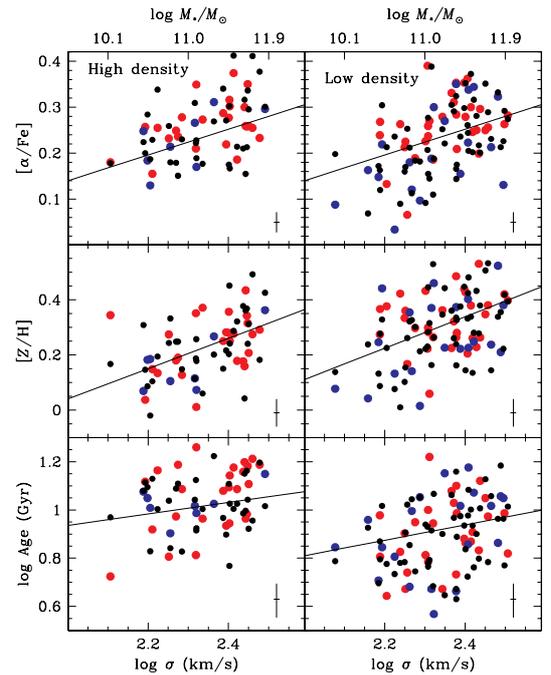}
\caption{From \citet{Thomas_etal_2005}. Stellar population parameters as a
  function of velocity dispersion and stellar mass for early type galaxies (red
  symbols are ellipticals, and blue symbols for lenticular). See original paper
  for details.}
\label{fig:fig6_thomas}
\end{center}
\end{figure}

Important constraints on the star formation history of galaxies can be obtained
from measurements of $\alpha$ element abundances. $\alpha$ elements are mainly
produced by supernovae Type II that explode on relatively short time-scales (of
the order of a few Myrs), whereas the iron peak elements are mainly produced by
supernovae Type Ia, on much longer time-scales.  Therefore, the [$\alpha$/Fe]
ratio can be used as an indicator of the star formation time-scale. Since the
early 90s, it has been known that brighter elliptical galaxies host stellar
populations that are $\alpha$ enhanced with respect to their lower mass
counterparts \citep{Worthey_Faber_Gonzalez_1992}. Fig.~\ref{fig:fig6_thomas}
shows stellar population parameters for early-type galaxies in different
environments, as a function of the galaxy velocity dispersion and stellar
mass. There is a clear trend for increasing [$\alpha$/Fe] ratios with
increasing stellar mass, both in low and high density regions. This trend
suggests that the time-scale of star formation correlates with the galaxy
stellar mass, and appears to be in contrast with naive expectations based on
the growth of dark matter haloes in hierarchical CDM cosmologies. Recent models
that take into account the suppression of late gas cooling (and hence star
formation) by radio-mode AGN feedback, however, have shown that the observed
behaviour is {\it not in contradiction} with the hierarchical paradigm of
structure formation \citep[e.g.][]{DeLucia_etal_2006}.  In these models, the
shape of the mass metallicity relation appears to be particularly sensitive to
the adopted feedback model
\citep{DeLucia_Kauffmann_White_2004,Bertone_DeLucia_Thomas_2007}, confirming
that these observations provide strong constraints on how and when baryons and
metals were ejected from galaxies.

\section{Metallicity distributions in the intra-cluster medium}

Simulations have shown that the baryon fraction in a rich cluster does not
evolve appreciably during its evolution. Clusters of galaxies can thus be
considered as closed systems, retaining all information about their past
star-formation and metal production histories. This suggests that direct
observations of element abundances in the intra-cluster medium (ICM) can
constrain the history of star formation in clusters, the efficiency with which
gas was converted into stars, the relative importance of different types of
supernovae, and the mechanisms responsible for the ejection and the transport
of metals. For all elements synthesised in stars after the primordial
nucleosynthesis, the energies corresponding to their K- and L-shell transitions
are accessible to modern X-ray telescopes. Most of the observed emission lines
can be converted to elemental abundances under the (reasonable) assumption of
collisional equilibrium. Complications due to extinction, non-equilibrium
ionization, optical depth effects, etc. are not significant, while the abundance
measurements depend crucially on the correct modelling of the temperature
structure \citep[][and references therein]{Werner_etal_2008}.

\begin{figure}[ht]
\begin{center}
\includegraphics[scale=0.35, angle=0]{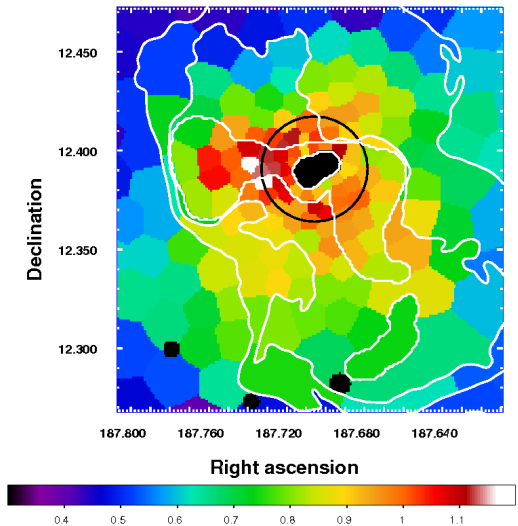}
\caption{From \citet{Simionescu_etal_2008}. Map of the iron abundance in
  M87. See original paper for details.}
\label{fig:fig10_simionescu}
\end{center}
\end{figure}

X-ray observations have shown that the ICM contains a considerable amount of
heavy elements, resulting in typical values of the order of one third solar or
even higher. Since a large fraction ($\sim 15-20$ per cent) of the total mass
of a cluster is in the ICM, and the galaxies contribute for a much smaller
fraction ($\sim 3-5$ per cent), assuming that the typical metallicity of
  cluster galaxies is about solar \citep{Renzini_1997} implies that there is
more mass in metals in the ICM than in all the galaxies in a cluster. Recent
observations have also shown that the typical metallicity in the ICM appears to
decrease by about 50 per cent to $z\sim 1$ \citep{Balestra_etal_2007}. Cool
core clusters (with peaked X-ray emission at their centres due to condensed
regions of cooler gas) usually exhibit centrally peaked metallicity
distributions, while flat distributions of metals are usually observed for non
cool-core clusters.  Observations of element abundances are consistent with the
idea that the excess of metals in the central regions of cool core clusters is
produced by supernovae Ia associated with the central cD galaxy
\citep{Tamura_etal_2001,DeGrandi_etal_2004}. If metals originated from the
stars produced in the central galaxy, the abundance profile would follow the
light profile in absence of mixing. The observed metal profiles are, however,
shallower than the corresponding light profiles, suggesting that metals are
mixed into the ICM and transported towards the outer regions by some physical
process(es).

Deep observations of bright clusters with XMM-Newton and Chandra allow the
construction of maps of the 2D distribution of metals in the ICM. Used together
with maps of thermodynamic properties, these maps can provide important
information about the merging history of the
cluster. Fig.~\ref{fig:fig10_simionescu} shows the map of Fe abundance in
M87. The half light radius of M87 is shown as a black circle while the 90~cm
radio emission contours are shown in white. In addition to a radial gradient,
which is usually observed in the central regions of galaxy clusters, the map in
Fig.~\ref{fig:fig10_simionescu} suggests the presence of deviations associated
with the inner radio lobes. These deviations are interpreted by
\citet{Simionescu_etal_2008} as a result of the influence of the central AGN on
the spatial distribution and transport of metals.

\begin{figure}[ht]
\begin{center}
\includegraphics[scale=0.3, angle=90]{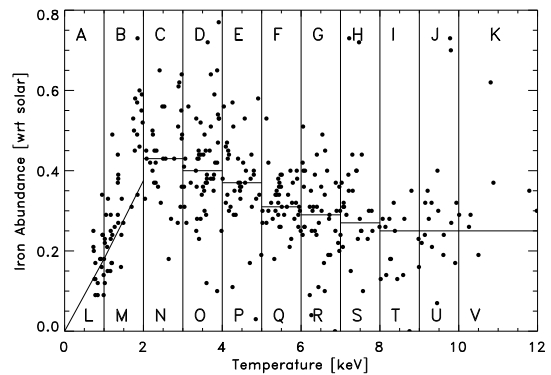}
\caption{From \citet{Baumgartner_etal_2005}. Iron abundance (with respect to
  solar) as a function of the ICM temperature. Each point is a single cluster
  measurement, while the lines show where the boundaries were placed for
  stacking individual measurements. See the original paper for details.}
\label{fig:fig2_baumgartner}
\end{center}
\end{figure}

Various additional physical processes can provide viable mechanisms for the
transfer of metals from the galaxies into the ICM, and several theoretical
studies have been carried out in order to quantify their relative importance
(for a recent review, see \citealt{Schindler_Diaferio_2008}). One physical
process that is expected to be more important in clusters than in average
regions of the Universe is {\it ram-pressure stripping}: galaxies travelling
through a dense intra-cluster medium suffer a strong pressure that sweeps cold
gas (and the metals in it) out of the stellar disc. Hydrodynamical simulations,
however, suggest that this process can account only for $\sim 10$ per cent of
the overall observed level of enrichment in the ICM within a radius of 1.3 Mpc
\citep{Domainko_etal_2006}. As noted by \citet{Renzini_1997}, there is also a
direct observational evidence that ram-pressure cannot play a dominant role in
the chemical enrichment of the ICM because this process would operate more
efficiently in high velocity dispersion clusters. A positive correlation
between the richness of the cluster and its metal content is not supported by
observations, as shown in Fig.~\ref{fig:fig2_baumgartner}.

Another obvious source of chemical enrichment is provided by supernova driven
winds, which are expected to play a dominant role in the outer regions of
galaxy clusters, where they can travel to larger distances due to the lower ICM
pressure. On the observational side, evidence in support of the existence of
outflows from galaxies has grown rapidly in the last years. Powerful galactic
winds appear to be ubiquitous in local starburst galaxies and in high-redshift
Lyman Break galaxies \citep{Heckman_2002}. 

Different methods to estimate the outflow rate suggest that this is of the same
order of magnitude of the star formation rate, with outflow speeds varying in
the range 400-800 km/s, and approximately independent of the galaxy circular
velocity. Observations also indicate, however, that galactic winds are complex
and multi-phase outflows of cool, warm, and hot gas, dust, and magnetized
relativistic plasma \citep{Veilleux_etal_2005}.  In addition, it is important
to remember that observational estimates of outflow rates should not be
confused with the rates at which mass and metals {\it escape} from galaxies,
because they are derived from material that is still quite deep within the
gravitational potential of the galaxy's dark matter halo. Whether the gas
reheated by supernova explosions will leave the halo depends on a number of
factors like the amount of intervening gas, and the fraction of energy lost by
radiative processes. Given these uncertainties and the dynamical complexity of
the outflows, an accurate implementation of the feedback process in
hydrodynamical simulations is still very difficult, and needs to rely on
`sub-grid' physics (i.e. this physical process is not modelled from first
principles, but `implemented' using theoretically and/or observationally
motivated relations, as is done in semi-analytic models).

\begin{figure}[ht]
\begin{center}
\includegraphics[scale=0.4, angle=-90]{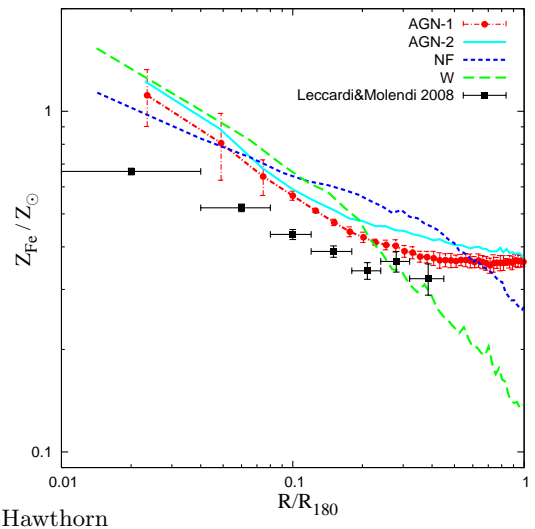}
\caption{From \citet{Fabjan_etal_2009}. Comparison between the observed and
  simulated profiles of emission-weighted Iron metallicity. Black symbols with
  error bars show observational measurements for galaxy clusters with
  $T_{500}>3$~kev. Different lines correspond to different simulation runs: no
  feedback (blue dashed), galactic winds (green dashed), and different AGN
  feedback implementations (red dot-dashed and cyan solid). See the original
  paper for details.}
\label{fig:fig8_fabjan}
\end{center}
\end{figure}

As noted in a recent review by \citet{Borgani_etal_2008}, hydrodynamical
simulations are in qualitative agreement with most recent observations. Some
problems are, however, persistent: e.g. too much gas cools at the centre of
massive clusters, leading to abundance profiles that are usually steeper than
observed. The problem is alleviated by the inclusion of AGN feedback:
Fig.~\ref{fig:fig8_fabjan} shows a comparison between observational data and
simulations including supernovae driven winds (dashed green line) and different
implementations of AGN feedback (red and cyan lines). In these simulations,
profiles from different runs have similar central slopes. This happens because
the effect that different feedback mechanisms have in displacing enriched gas
and regulating star formation almost balance each other in the central
regions. In the outer regions (at cluster-centric distances $\gtrsim
0.2\,R_{200}$), the metal abundance profile in runs including AGN feedback (red
and cyan lines) becomes flatter with respect to the profile obtained in a run
with only galactic winds (green line), resulting in a better agreement with
observational measurements. The flattening in the outer regions is due to the
fact that AGN feedback displaces large amounts of enriched gas from star
forming regions at high redshift and, at the same time, efficiently suppresses
cooling at low redshift.

\section{Metallicity distributions in the inter-galactic medium}
\label{sec:igm}

The existence of metals in the inter-galactic medium (IGM) has been known for
about two decades by their absorption lines in the spectra of high-redshift
quasars.  First indications by \citet{Meyer_York_1987} were later confirmed
with the advent of 8m-class telescopes and high-resolution spectra, that
clearly showed the presence of measurable amounts of heavy elements (about one
per cent solar) in the Lyman alpha forest absorbers
\citep[e.g.][]{Songaila_Cowie_1996}. More recent studies have indicated that
heavy elements are present at all densities, down to at least the mean cosmic
density \citep{Schaye_etal_2003} and that, on small scales, the distribution of
metals in the IGM is highly inhomogeneous. 

Among the different types of QSO absorbers, damped Lyman alpha systems (DLAs)
have received particular attention (for a review, see
\citealt{Wolfe_Gawiser_Prochaska_2005}). These objects are characterised by HI
column densities $N(HI) \gtrsim 2\times 10^{20}\,{\rm cm}^{-1}$, and hydrogen
appears to be mainly neutral in these systems, implying that they represent
important neutral gas reservoir for star formation at high redshift.  A nice
review of recent metallicity measurements in DLAs is provided by
\citet{Pettini_2006}. Data show that the mean metallicity of these systems is
relatively low (about one tenth solar or lower), but that individual
measurements span a wide range of metallicities, from solar to less than one
hundredth solar. A few measurements of element abundance patters are available,
and suggest that, at least some DLAs, have [$\alpha$/Fe] ratios lower than those
measured in Milky Way stars of low metallicities. This has been interpreted as
an indication of low levels of star formation
\citep[e.g.][]{Matteucci_Recchi_2001}.

\begin{figure}[ht]
\begin{center}
\includegraphics[scale=0.35]{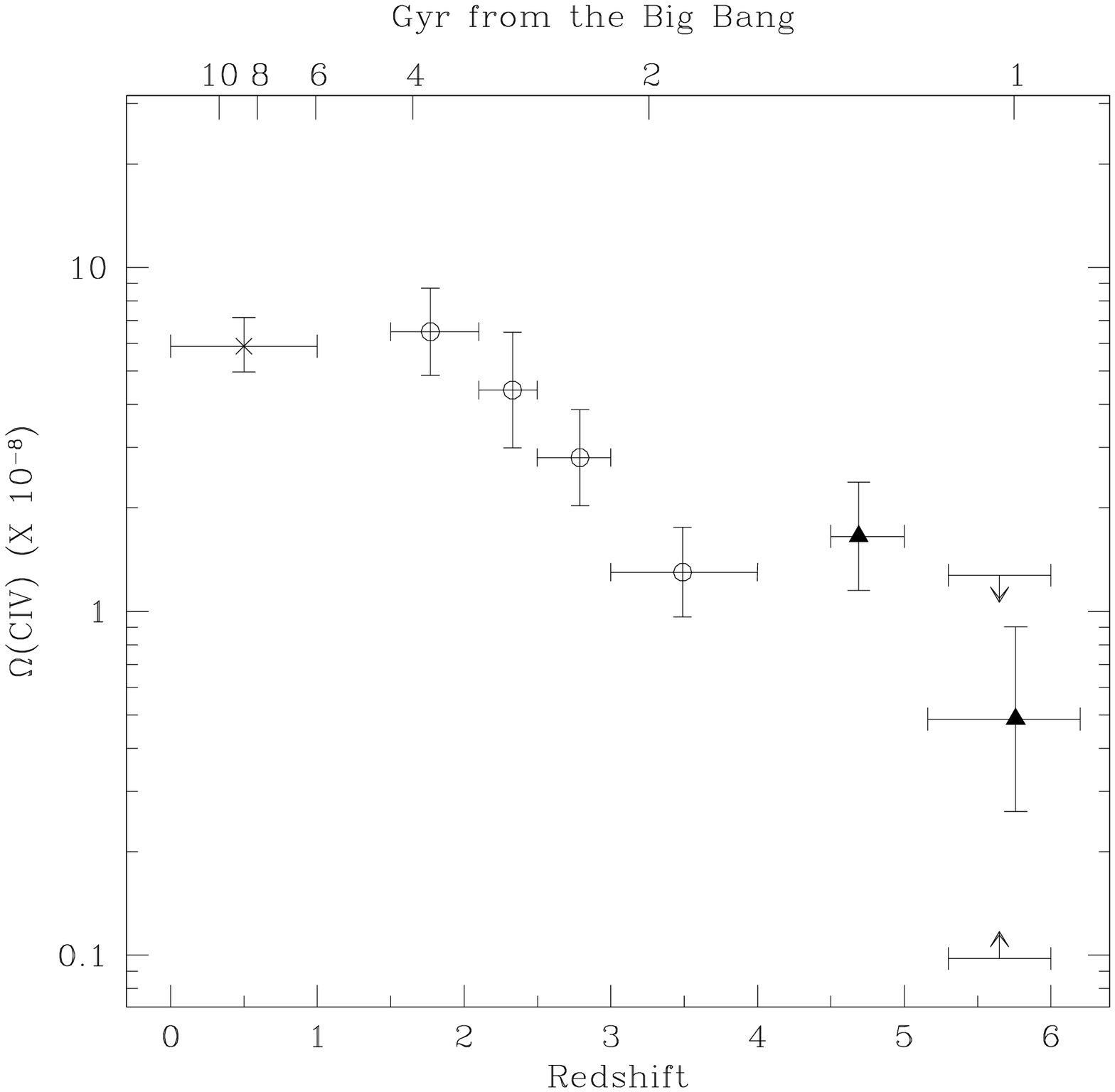}
\caption{From \citet{Dodorico_etal_2009}. Mass density of C IV as a function of
  redshift. The lowest redshift data point is from \citealt{Cooksey_etal_2009},
  solid triangles are from \citealt{Pettini_etal_2003} and
  \citealt{Ryan-Weber_etal_2009}. Finally, the 95 per cent confidence interval
  at the highest redshift considered are from \citealt{Becker_etal_2009}.}
\label{fig:fig7_dodorico09}
\end{center}
\end{figure}

Several recent studies have tried to constrain the cosmic evolution of the
metal content of the IGM through measurements of the CIV mass density,
$\Omega_{\rm CIV}$. Fig.~\ref{fig:fig7_dodorico09} is from a recent study by
\citealt{Dodorico_etal_2009} and shows a significant increase of $\Omega_{\rm
  CIV}$ at $z \lesssim 3$ and a downturn at $z >5$
\citep{Becker_etal_2009,Ryan-Weber_etal_2009}. A satisfactory theoretical
interpretation for the trends shown in Fig.~\ref{fig:fig7_dodorico09} is still
missing. The most recent theoretical study in this area is that by
\citealt{Oppenheimer_Dave_2008}. These authors have used cosmological
simulations which include feedback in the form of momentum driven winds, and
found an apparent lack of evolution of the CIV mass density from $z\sim 5$ to
$z\sim 1.5$. In the simulations, this results from an overall increase in the
cosmic abundance of carbon, and a parallel reduction of the fraction of triply
ionized carbon. None of the simulations presented in this study predict the
recently observed increase of $\Omega_{\rm CIV}$ towards lower redshift.

From the theoretical viewpoint, the details of how large volumes of the IGM
could have been seeded by the products of stellar nucleosynthesis at early
times, and how these products could reach such low overdensities remain to be
understood. Galactic winds might play an important role in polluting the IGM,
particularly at high redshift, when the star formation rates are more elevated
and the energy deposited by supernovae can potentially disrupt the shallow
potential wells of proto-galaxies. The observed enhancement of metal abundances
in the proximity of Lyman-break galaxies suggests that star forming galaxies
might be an important source of later chemical enrichment. A number of recent
numerical studies have focused on comparing different feedback schemes with
various observational measurements
\citep[e.g.][]{Oppenheimer_Dave_2008,Tescari_etal_2009}. These studies confirm
that galactic winds are indeed able to enrich large comoving volumes of the IGM
at high redshift. Numerical results are in qualitatively good agreement with
observational measurements, but not without problems. For example, in these
simulations the metallicity distributions of DLAs (as traced by the velocity
widths) and the observed recent increase of $\Omega_{\rm CIV}$ (as mentioned
above) are not well reproduced, suggesting that either the feedback scheme
adopted is too simplistic/incorrect, or that there are some physical mechanisms
not included in the simulations (e.g. enrichment by Population III stars,
turbulence, etc.).

\section{Metal abundances at high redshift}

\begin{figure}[ht]
\begin{center}
\includegraphics[scale=0.4, angle=0]{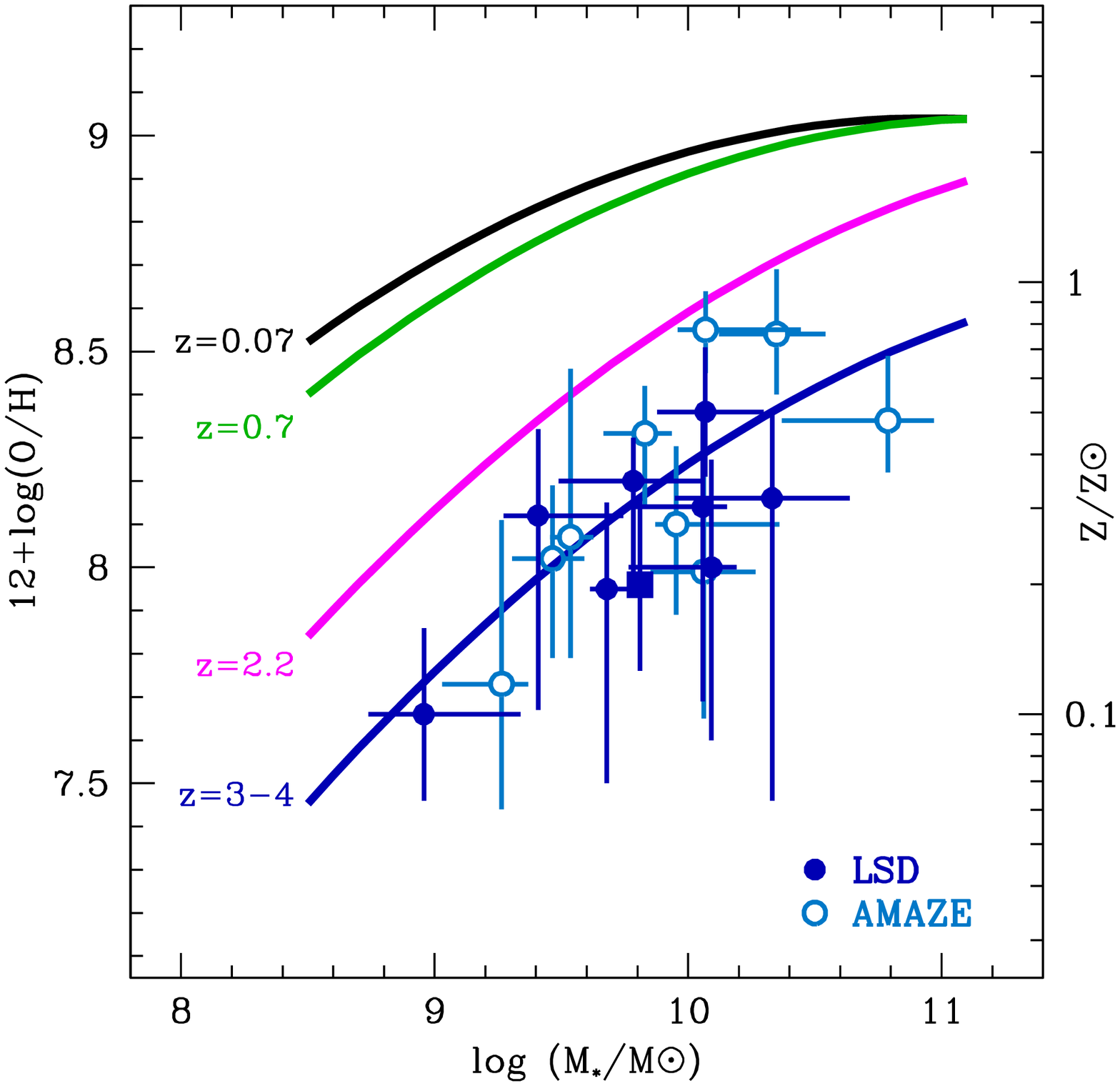}
\caption{From \citet{Mannucci_etal_2009}. Evolution of the mass-metallicity
  relation from z =0.07 \citep{Kewley_Ellison_2008}, to z =0.7
  \citep{Savaglio_etal_2005}, z =2.2 \citep{Erb_etal_2006}, and $z =3-4$
  \citep{Maiolino_etal_2008,Mannucci_etal_2009}. All data have been calibrated
  to the same metallicity scale and IMF. The lines show quadratic fits to the
  data (see original papers for details).}
\label{fig:fig5_mannucci}
\end{center}
\end{figure}

Several studies have recently focused on the evolution of the mass-metallicity
relation as a function of redshift. Results are summarised in
Fig.~\ref{fig:fig5_mannucci}. As discussed earlier, the shape of the
mass-metallicity relation provides important constraints on feedback mechanisms
and on their efficiency as a function of the galaxy's potential
well. Particular care, however, needs to be taken when interpreting the data
shown in Fig.~\ref{fig:fig5_mannucci}. At different redshifts, different
emission lines are available, so that different calibrations need to be
used. The strong effects on the shape and zero-point of the metallicity
calibration can lead to incorrect evolutionary interpretations. In addition,
samples observed at different epochs do represent different classes of objects
that are not necessarily linked from an evolutionary point of view. This also
complicates any comparison with galaxy formation models and/or simulation
results.

\begin{figure}[ht]
\begin{center}
\includegraphics[scale=0.36, angle=-90]{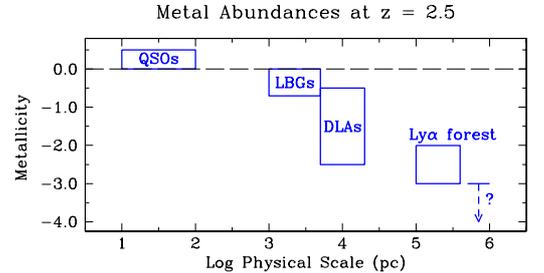}
\caption{From \citet{Pettini_2006}. Snapshot of the metallicity of different
  components of the high redshift universe. The logarithm of the metallicity is
  plotted relative to solar (indicated by the long-dash line at 0.0) as a
  function of the typical linear scale of the structure to which it refers.}
\label{fig:fig11_pettini}
\end{center}
\end{figure}

A (now incomplete) summary of our current knowledge about element abundances at
$z\sim 2.5$ is provided in Fig.~\ref{fig:fig11_pettini} from
\citet{Pettini_2006}. This figure shows the metallicities measured in different
systems at high redshift, as a function of the typical linear scale of the
structure to which it refers. The figure shows a clear trend of decreasing
abundances with increasing scale. The gas in the inner 10-100 pc of the highest
overdensities, where supermassive black holes reside, is enriched to solar or
supersolar levels. Actively star-forming galaxies sample physical scales of the
order of a few kpc, and are enriched to near solar metallicity. DLAs likely
sample more diffuse gas at the outskirts of galaxies in the process of
formation and, as mentioned earlier, they represent a quite heterogeneous
population with a large scatter in the degree of chemical enrichment. Finally,
the Lyman $\alpha$ forest is believed to trace large scale structures of
moderate overdensity relative to the cosmic mean, and it appears to contain
only trace amounts of metals. Fig.~\ref{fig:fig11_pettini} shows that, at any
cosmic epoch, the degree of metal enrichment depends strongly on the scale
considered.

Integrating the comoving star formation rate deduced from galaxy surveys, it is
possible to calculate (adopting some conversion factors) an estimate of the
comoving density of metals which should have accumulated by a given
redshift. As discussed by \citet{Pettini_2006}, when adding together the
contributions shown in Fig.~\ref{fig:fig11_pettini}, one can account for about
$40$ per cent of the metal production from the Big Bang to $z = 2.5$. The
contribution from sub-DLAs (not included in the calculation by Pettini) amounts
to about 6 per cent of the expected metals at $z\sim 2.5$
\citep{Peroux_etal_2007}. The {\it missing metals problem} has been revisited
by \citet{Bouche_etal_2007} who estimated that currently known galaxy
populations at $z\sim 2$ can account for $\gtrsim 30$ to $\lesssim 60$ per cent
of the metals expected, which is approximately what is required to close the
metal budget. As noted by Bouch\'e and collaborators, however, it is unclear
what the overlap is between the various galaxy populations at this redshift. A
conservative estimate provided by these authors, is that we can currently
account for $\gtrsim 65$ per cent of the metals expected at $z\sim 2.5$, the
metals being equally spread between galaxies and IGM.

Still uncertain is the contribution provided by the so-called Warm Hot
Intergalactic Medium (WHIM), a tenuous and elusive phase that is predicted by
numerical simulations. This baryonic phase is expected to be characterised by
relatively high temperatures (T$\sim10^5-10^7$ Kelvin), and is predicted to
radiate most of its energy in the ultraviolet and X-ray bands. Current X-ray
telescopes do not have enough sensitivity to detect the hotter WHIM, but future
space missions (e.g. Constellation X, XEUS) should be able to detect both
emission lines and absorption systems from highly ionised atoms. Interestingly,
recent numerical simulations show that the metal distribution in this phase
depends significantly on the details of feedback \citep{Tornatore_etal_2009},
therefore providing an additional tool to study the physical processes
responsible for the chemical enrichment of the Universe.

\section{Discussion and conclusions}

As discussed above, metals are widespread in our Universe and different
techniques have to be used to measure their amounts in different baryonic
components (stars, cold inter-stellar medium, hot intra-cluster gas,
inter-galactic medium). At least for some of these techniques, there are still
large uncertainties on the observational calibration, which complicates the
comparison between measurements at different redshift and/or in different
bands. 

\begin{figure*}[ht]
\begin{center}
\includegraphics[scale=0.65]{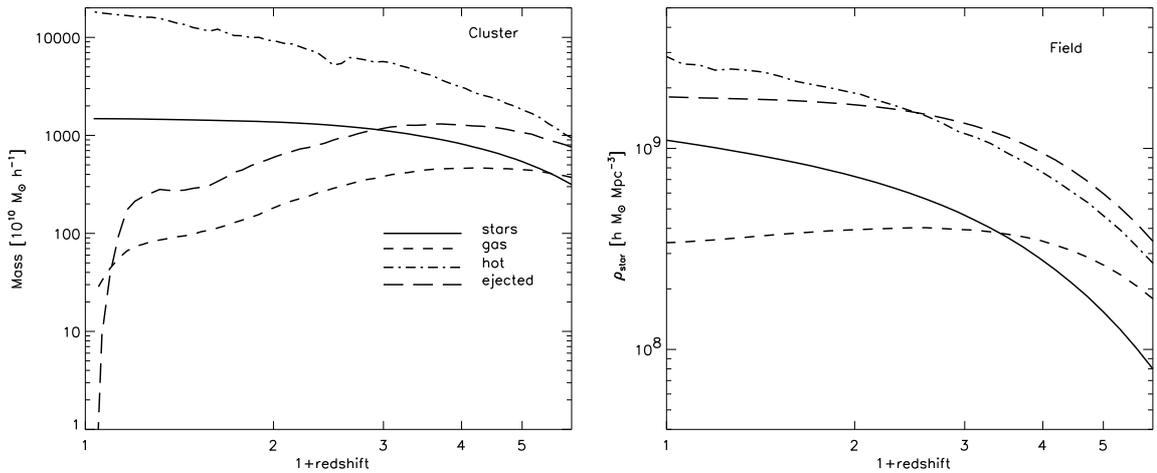}
\caption{From De Lucia et al. (2004). Evolution of the metal content in
  different phases. The solid line represents the evolution of the metal
  content in stars, the dashed line the cold gas, the dashed-dotted line the
  hot gas, and the long-dashed line the ejected component. The metal content in
  each phase is normalised to the total mass of metals produced from all
  galaxies considered. The left panel shows the results for galaxies within the
  virial radius of a massive ($\sim 10^{15}\,{\rm M}_{\odot}$) cluster, while
  the right panel shows the results for a typical `field' region. See original
  paper for details.}
\label{fig:fig13_pap2004}
\end{center}
\end{figure*}

Much work is going on to improve the precision of the observational
measurements and to push element abundance studies to higher and higher
redshift. Stellar population synthesis models have been improved significantly
in recent years: many new empirical libraries have been made available with
improved spectral resolution and parameter coverage
\citep[e.g.][]{Sanchez_Blazquez_etal_2006}, and many libraries of theoretical
high-resolution stellar spectra for both scaled-solar and $\alpha$ enhanced
chemical mixtures have been published \citep[][and references
  therein]{Coelho_etal_2007}. Parallel effort is going in the direction of
refining photoionization models (such as MAPPINGS - \citealt{Dopita_etal_2005})
which help us correctly diagnose physical properties of star forming
galaxies. Accurate measurements of metallicities for much larger samples of
individual stars for our Galaxy and its satellites will be available in the
coming years from a number of ongoing astrometric and spectroscopic surveys
(e.g. the satellite {\it Gaia} - \citealt{Perryman_etal_2001}; the Sloan
Extension for Galactic Understanding and Exploration -
\citealt{Beers_etal_2004}). Future X-ray satellites will likely push the study
of the distribution of metals in the ICM to larger radii in nearby clusters,
and allow metallicity profiles to be measured at higher redshift \citep[see
  e.g.][]{Giacconi_etal_2009}. High-resolution spectroscopy in the X-ray and in
the UV (e.g. with the Cosmic Origin Spectrograph onboard the Hubble Space
Telescope) will also allow us to make progress in the study of the diffuse
warm-hot baryons in the local Universe.

From the theoretical viewpoint, hydrodynamical simulations and
semi-analytic models provide the interpretative framework for these
observational studies. Many theoretical predictions are already available but
not always these studies have been able to provide a `self-consistent'
description of the circulation of the baryons in different components. Indeed,
cosmological hydrodynamical simulations have mostly focused on the distribution
of metals in the IGM and ICM, but the numerical resolution adopted is
insufficient to study the metal distribution in galaxies. On the other hand,
semi-analytic models can provide detailed predictions on the distribution of
metals in the stellar and gaseous components but do not provide details on the
spatial distribution of the metals.

Fig.~\ref{fig:fig13_pap2004} shows the evolution of the metal content in
different phases for one particular feedback scheme implemented in a
semi-analytic code by De Lucia et al. (2004). The left panel corresponds to a
cluster simulation while the right panel is for a region with average
density. Comparison between the two panels shows that cosmic variance effects
are going to play an important role when trying to constrain different feedback
schemes using observational metallicity measurements. Nevertheless, the same
study pointed out (and many later studies confirmed) that the fraction of
metals (as well as their distribution) in different baryonic components varies
significantly when different feedback schemes are adopted. It is therefore to
be expected that the vast amount of data that will be available in the next
future, and detailed comparisons with theoretical models will eventually tell us
a great deal about how galaxies ejected their metals over the history of the
Universe.

\section*{Acknowledgments} 
I thank the organisers of the conference ``Galaxy Metabolism: Galaxy Evolution
near and far'' (Sydney, 22-26 June 2009) for financial support which helped me
take part in this very stimulating meeting. I acknowledge financial support
from the European Research Council under the European Community's Seventh
Framework Programme (FP7/2007-2013)/ERC grant agreement n. 202781. It is a
pleasure to thank Stefano Borgani, Francesca Matteucci, Luca Tornatore and
Paolo Tozzi for many stimulating discussions about chemical enrichment, and for
useful comments on a preliminary version of this manuscript. I am grateful to
Anna Gallazzi and Christy Tremonti for many clarifications about observational
data analysis, and many useful discussions.

%\end{multicols}

\end{document}